# Voltage-Induced Oxidation for Enhanced Purity and Reproducibility of Quantum Emission in Monolayer 2D Materials

Sung-Joon Lee[a][†], Hsun-Jen Chuang*[†], Kathleen M. McCreary, Mehmet A. Noyan, Berend T. Jonker

U.S. Naval Research Laboratory, Washington, DC, 20375, USA

[a]Postdoctoral associate at the U.S. Naval Research Laboratory through the American Society for Engineering Education

[†]These authors contributed equally to this work.

*Email: hsunjen.chuang.civ@us.navy.mil

**ABSTRACT**

We report a voltage-induced oxidation technique using conductive atomic force microscopy to enhance the single-photon purity and reproducibility of quantum emitters in monolayer tungsten diselenide ($WSe_2$). By applying a controlled electric field across a monolayer $WSe_2$/poly(vinylidene fluoride-co-trifluoroethylene) (P(VDF-TrFE)) on a silicon substrate, localized oxidation is induced around nanoindented emitter sites in the $WSe_2$. This treatment selectively suppresses defect-bound exciton emissions while preserving emission from pristine regions within the indentations. Photoluminescence and second-order correlation measurements at 18 K demonstrate a substantial increase in single-photon purity when comparing emitters from untreated and voltage-treated regions. Emitters from untreated regions showed average values of $g^2(0)$ near or above the 0.5 threshold. In contrast, emitters from voltage-treated regions exhibited $g^2(0)$ values consistently below 0.14, with most falling near 0.05, demonstrating high-purity single-photon emission well below the $g^2(0) < 0.5$ threshold. This enhancement results from the oxidation-induced suppression of spurious luminescence from the area around the quantum emitter site that is spectrally degenerate with the single-photon wavelength. This approach offers nonvolatile, spatially selective control over the emitter environment without degrading the emission intensity, improving both purity and stability. It provides a scalable route for integrating high-quality quantum emitters in two-dimensional materials into photonic platforms. Integration with spectral tuning strategies such as strain engineering, local dielectric patterning, or electrostatic gating could further enable deterministic, wavelength-selective single-photon sources for advanced quantum photonic applications.

**INTRODUCTION**

Two-dimensional (2D) materials, including van der Waals heterostructures,[1–3] have become essential platforms for probing emergent quantum phenomena such as spintronic behavior and multiferroicity.[4] Recent advancements in the synthesis and integration of high-quality 2D materials further underscore their potential in next-generation electronic and quantum devices.[5–7] These quantum effects not only deepen our understanding of condensed matter systems but also enable the development of advanced technologies, including high-performance transistors, neuromorphic computing architectures, and quantum information processing.[8–11]

Among these, quantum emitters (QEs), or single-photon emitters (SPEs), embedded in monolayer 2D materials have attracted considerable interest due to their critical roles in quantum photonic circuits, secure communication, and precision sensing.[9,10] As quantum photonic technologies aim to surpass the limitations of classical optical systems, developing QEs with high single-photon purity, deterministic spatial control, and reproducible positioning is imperative.

While significant research efforts have focused on conventional three-dimensional (3D) materials such as diamond and III-V quantum dots,[9,10] these systems face substantial challenges, including complex fabrication processes and integration constraints. In contrast, 2D materials, such as monolayer transition metal dichalcogenides (TMDs), offer distinct advantages for QE development due to their atomic-scale thickness, which enables seamless integration onto various substrates without lattice matching constraints. Furthermore, their atomically thin nature enhances photon extraction efficiency, making TMD-based QEs particularly promising for photonic integrated circuits (PICs) and waveguide architectures.

Various techniques have been explored to create deterministic QEs in TMDs, including nanoparticle-assisted localization,[12] substrate-induced strain,[13] nanopillar engineering,[14–17] and nanoindentation.[18,19] Despite these advancements, challenges persist in achieving consistently high-yield and high-purity QEs. Primary factors limiting the purity of QEs include: (1) spectral overlap from closely spaced quantum emissions due to non-uniform strain fields and complex defect states around the indentation sites,[20] and (2) unwanted semi-classical background emission from defect-bound excitons. These defect-bound excitons emit semi-classical light that is spectrally degenerate with the emission wavelength of QEs, complicating spectral isolation. Additionally, the diffraction-limited nature of light collection in these systems further dilutes the measured single-photon purity by capturing unwanted emissions beyond the nanometer-scale QE sites. Consequently, precise control of emission wavelength and reliable suppression of background emissions remain critical challenges, impeding scalable quantum photonic integration of these QEs.

In previous work, we achieved enhanced single-photon purity in monolayer WSe$_2$ by applying gate voltages to induce electric-field-assisted dissociation of defect-bound excitons.[21] Although effective, the improvement was reversible and required continuous external voltage application, highlighting the need for nonvolatile methods to achieve stable high-purity emission. Subsequently, we demonstrated nonvolatile, reversible control of QE purity in monolayer tungsten disulfide (WS$_2$) through ferroelectric polymer integration using poly(vinylidene fluoride-co-trifluoroethylene) (P(VDF-TrFE)).[22] By switching the ferroelectric polarization state of P(VDF-TrFE) with applied voltages, we effectively toggled emission purity between quantum and semi-classical regimes. Additionally, leveraging strong charge transfer effects between conductive materials (such as graphite) and WSe$_2$ enabled substantial suppression of defect-bound exciton emissions, providing another promising route for achieving nonvolatile, high-purity quantum emission.[23] Nonetheless, the development of reliable and reproducible techniques for generating consistently high-purity QEs remains a continuing challenge.

In this study, we introduce a voltage-induced oxidation (VIO) treatment utilizing conductive atomic force microscopy (c-AFM) to significantly enhance single-photon purity and the reproducibility of QEs in monolayer tungsten diselenide (WSe$_2$). The controlled voltage treatment selectively oxidizes WSe$_2$ regions surrounding the QE sites, while preserving the emitters themselves. This targeted treatment suppresses unwanted classical emissions while maintaining QE integrity. This localized modification avoids the need for continuous external fields and enables nonvolatile operation, making it well-suited for scalable photonic device integration. This selective oxidation simplifies the photoluminescence (PL) spectrum by spectrally isolating QE emissions and substantially reducing background emission from defect-bound excitons. Single-photon purity $P = 1 - g^{(2)}(0)$ was quantified by measuring the second-order correlation function $g^{(2)}(t)$. A dramatic improvement in single-photon purity following VIO treatment is observed, increasing from an average of approximately 40% in untreated emitters to over 92% after treatment. Notably, the maximum achieved purity after treatment exceeded 97%, underscoring the effectiveness of the voltage-induced oxidation approach in suppressing background emission and isolating high-quality quantum emitters. Furthermore, the VIO approach enhances the reproducibility of high-purity QEs compared to previously reported methods described above. Although the ultimate purity may presently be limited by initial material quality, our results demonstrate substantial improvements achievable through optimized monolayer preparation. This VIO approach thus offers a robust, scalable, and efficient strategy for realizing high-yield, high-purity QEs from 2D materials, paving the way toward practical implementations of QEs in quantum photonic applications.

## RESULTS AND DISCUSSION

Figures 1(**a**) and 1(**d**) illustrate the fabrication process of high-yield, high-purity QEs in monolayer WSe$_2$ via VIO treatment. The sample structure consists of monolayer CVD-grown WSe$_2$ transferred onto a ~260 nm-thick P(VDF-TrFE) polymer film, which itself is transferred onto a Si substrate. Transferring (rather than spin-coating) the P(VDF-TrFE) film provides a smoother and more uniform surface,[22] which not only facilitates consistent nanoindentation but also ensures more reliable transfer of monolayer WSe$_2$. The thickness was selected based on prior experimental results showing that nanoindentation to a depth of ~200 nm yields a high density of QEs.[22] Graphite flakes are then transferred to partially cover the edge of the WSe$_2$, enabling lateral electrical contact without covering the indentation sites. An array of indentations is subsequently created in the monolayer WSe$_2$ using AFM nanoindentation,[18] as confirmed by the optical microscopy (OM) image in Figure 1(**b**), which was captured at room temperature before voltage treatment.

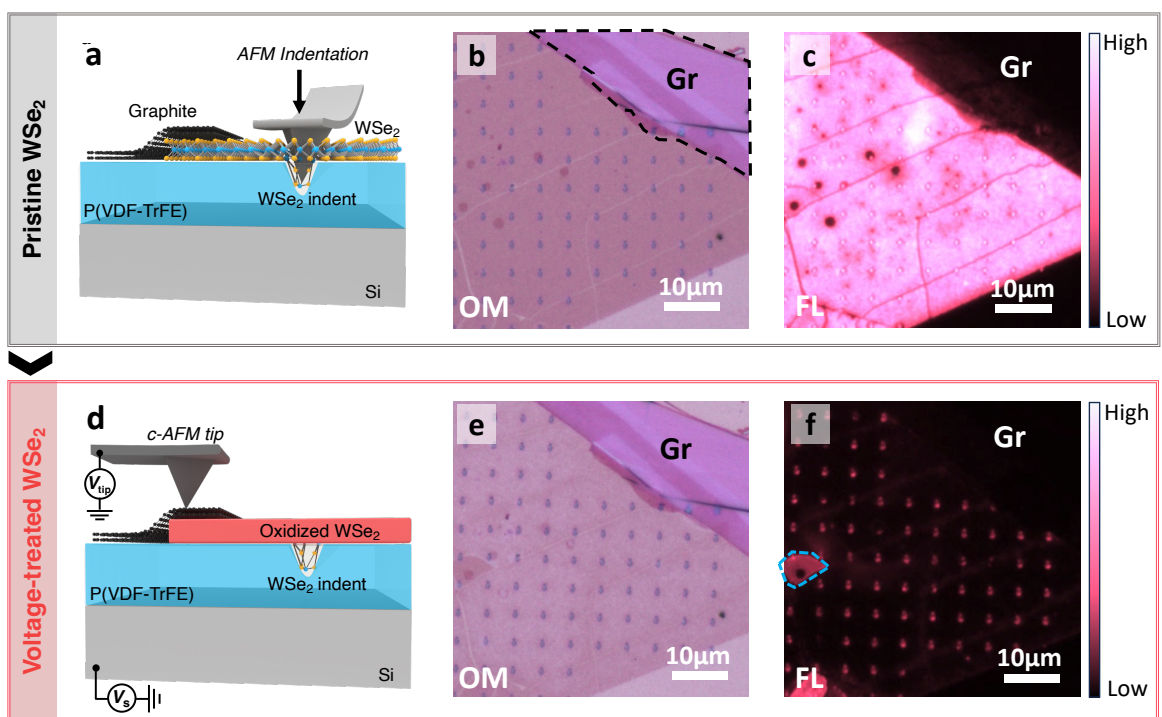

**Figure 1. Schematic and optical characterization of monolayer WSe$_2$ quantum emitters before and after voltage-induced oxidation (VIO).** (**a**) Schematic illustration of quantum emitter (QE) formation via AFM nanoindentation on monolayer CVD-WSe$_2$/P(VDF-TrFE)/Si. (**b,c**) Optical microscopy (OM) and fluorescence (FL) images, respectively, showing pristine WSe$_2$ before voltage treatment. (**d**) Schematic of the VIO process, where a controlled bias is applied via graphite top contact and a conductive AFM (c-AFM) tip to induce oxidation of the WSe$_2$ outside the indentation. (**e,f**) OM and FL images after VIO treatment, respectively. A clear material contrast change is observed in (e), accompanied by strong suppression of FL intensity in (f) suggesting preservation of pristine WSe$_2$ at the indentation sites and effective oxidation of the surrounding material. Blue dashed lines in (f) mark regions of persistent fluorescence attributed to cracks in the WSe$_2$ layer that locally inhibited oxidation. OM and FL images were collected under identical contrast/illumination settings to enable direct visual comparison. All images were taken at room temperature.

The P(VDF-TrFE) polymer was selected because it appears to promote oxidation of $WSe_2$ under voltage treatment more effectively than non-polar polymers such as poly(methyl methacrylate) (PMMA). This enhanced reactivity may stem from interfacial effects such as charge transfer or local polarization-induced p-type doping in the $WSe_2$ layer, as suggested by previous studies showing ferroelectric-field-driven modulation of carrier type and optical properties in TMD/P(VDF-TrFE) heterostructures.[24,25] This property is likely influenced by surface effects such as charge exchange interactions or p-type doping of $WSe_2$ by P(VDF-TrFE), which enhances current flow between the $WSe_2$ and the polymer.[22] These factors collectively make P(VDF-TrFE) an effective medium for enabling localized voltage-induced oxidation and for stabilizing high-yield quantum emitter formation.

The fluorescence (FL) microscopy images in Figure 1(**c**) reveal bright emission spots coinciding with the indentations in Figure 1(**b**), confirming that nanoindentation effectively preserves pristine $WSe_2$ within indentation sites. After applying a controlled voltage through graphite by c-AFM tip, selective oxidation of $WSe_2$ occurs, as schematically illustrated in Figure 1(**d**). The OM and FL images after voltage treatment shown in Figures 1(**e**) and 1(**f**) respectively demonstrate clear contrast changes and significantly reduced FL intensity surrounding the indentation areas compared to the pre-treatment images in Figures 1(**b**) and 1(**c**). Importantly, FL intensity within the indentation sites remains bright as shown in Figure 1(**f**), indicating pristine $WSe_2$ preservation. Notice the regions within cracks labeled in blue dashed lines in Figure 1(**f**) show the bright FL intensity. We attribute to disrupted current flow which hinders complete oxidation of those areas.

These observations strongly suggest that global material modification, specifically oxidation of the monolayer $WSe_2$ has been induced by the applied voltage treatment. We attribute the onset of VIO in monolayer $WSe_2$ to electric fields exceeding a critical threshold. This proposed VIO is primarily driven by the strong electric field generated at the tip-sample junction. When a bias is applied to a conductive AFM tip in contact with a $WSe_2$ monolayer, it can induce localized oxidation, where the field enhances the transport of ionic species, (e.g., oxygen ions or hydroxyl groups derived from ambient humidity) toward the $WSe_2$ surface. This field can also lower the activation energy required for the oxidation reaction to occur at the atomic level.[26–28] Similarly, Kang *et al.* demonstrated that applying a local electric field via an AFM tip oxidizes the contact area of $WS_2$, confirming a field-assisted oxidation mechanism.[29] A stronger electric field may further accelerate the generation of reactive oxygen species at the tip-sample interface, promoting oxidation of the $WSe_2$.

In this study, rather than applying a bias through a conductive AFM tip directly in contact with $WSe_2$ to induce localized oxidation, we utilized a graphite flake placed on $WSe_2$ monolayer. The graphite was then contacted by the biased c-AFM tip. This Gr-$WSe_2$/P(VDF-TrFE)/Si capacitor configuration enables the generation of a uniform electric field and effectively initiates oxidation of the WSe$_2$. Applying a voltage of approximately 13 V across the ~260 nm-thick P(VDF-TrFE) layer corresponds to an electric field of ~0.5 MV cm$^{-1}$, which initiates WSe$_2$ oxidation. This effect was not observed with voltages below 10 V, indicating a clear field threshold for inducing material modification. These findings align with previous reports of field-induced breakdown and oxidation in monolayer TMDs such as MoS$_2$ and WSe$_2$, where irreversible transformations occur under high electric fields.[30–32]

Figure S1 in the Supporting information shows sequential optical images from the AFM camera highlighting the time-dependent oxidation of indented monolayer WSe$_2$. A total bias of 20 V is applied by ramping the tip voltage from 0 V to +10 V while the substrate is held constant at -10 V. The tip is then held at +10 V for several minutes, during which progressive OM contrast changes are evident in the optical images, indicating ongoing oxidation. These results confirm that both the voltage magnitude and exposure time are critical parameters for inducing oxidation. We note that complete oxidation, including within indentations, can occur with prolonged high-voltage application.

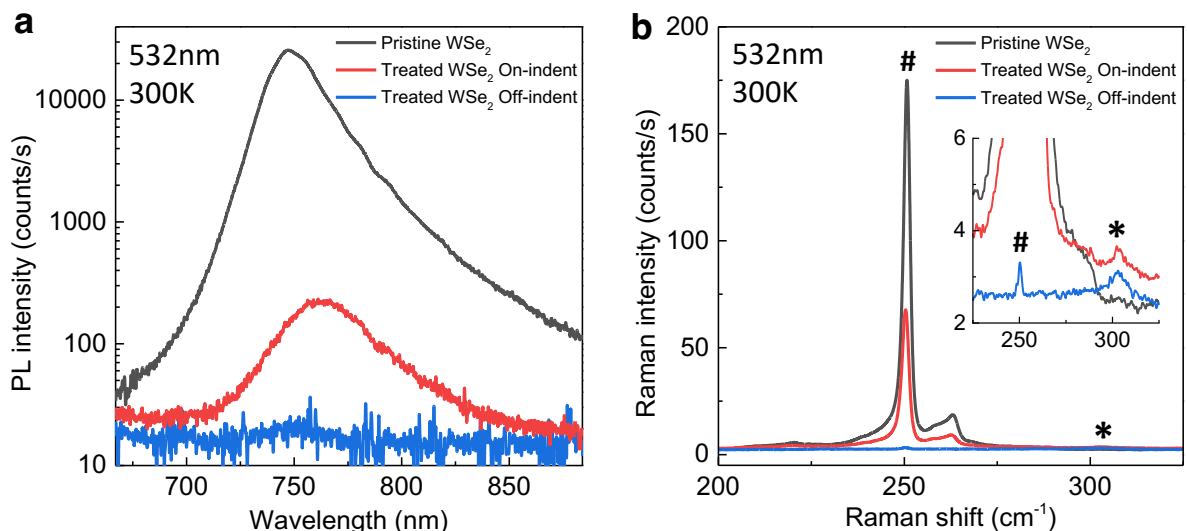

**Figure 2. Photoluminescence (PL) and Raman spectroscopy comparison of pristine and voltage-treated monolayer WSe$_2$.** (**a**) The pristine monolayer WSe$_2$ (black) exhibits a broad emission peak centered around 750 nm. The voltage-treated sample shows complete quenching of PL in off-indent regions (blue), indicating effective oxidation. In contrast, on-indent regions (red) retain visible PL emission with reduced intensity, consistent with spatially localized preservation of pristine WSe$_2$. (**b**) Raman spectra collected from the same regions as in (a). The predominant peak near ~250 cm$^{-1}$ (labeled "**#**") corresponds to the $A^1_g/E^1_{2g}$ vibrational modes of WSe$_2$. This mode is retained in both pristine and treated on-indent regions but is significantly diminished in off-indent regions, indicating structural degradation from oxidation. A new Raman mode near 300 cm$^{-1}$ (labeled "*****") appears in the off-indent (blue) region and partially in the on-indent (red) region, and is attributed to the O-W-O bending mode characteristic of oxidized WSe$_2$. The appearance of this oxidation related peak in both regions is again attributed to the finite laser spot size relative to the indentation dimensions. All PL and Raman measurements were conducted under CW 532 nm laser excitation.

Room-temperature photoluminescence (PL) and Raman spectroscopy with 532 nm CW excitation further validate material modification resulting from post-VIO treatment. Figure 2(**a**) compares PL spectra of pristine WSe$_2$ (black), voltage-treated WSe$_2$ on-indent (red), and voltage-treated WSe$_2$ off-indent (blue), with their corresponding measurement locations shown in Figure S2 in the Supporting Information. The pristine WSe$_2$ (black; untreated) exhibits a broad PL emission peak near 750 nm in good agreement with the literature. In contrast, voltage-treated WSe$_2$ off-indent regions (blue) show no detectable PL, indicating strong oxidation-induced quenching. On-indent treated regions maintain reduced but visible PL (red), consistent with the preservation of WSe$_2$ in indentation sites. The reduced PL intensity compared to pristine WSe$_2$ is attributed to the laser spot size (~2 μm) being much larger than the indentation diameter (~300-600 nm, see Figure S3), such that the PL signal from pristine WSe$_2$ is collected over a broader area, enhancing its apparent intensity. These results, together with the FL contrast from Figure 1(**c**) to Figure 1(**f**), further confirm that VIO modifies the surrounding WSe$_2$ while preserving emitters inside the indents.

Raman spectroscopy provides further evidence of structural modification resulting from VIO treatment. In Figure 2(**b**), we compare Raman spectra from pristine WSe$_2$ (black), voltage-treated WSe$_2$ on-indent (red), and voltage-treated WSe$_2$ off-indent (blue). The characteristic monolayer WSe$_2$ Raman modes ($A^1_g$ + $E^2_g$) around 250 cm$^{-1}$ with 532 nm CW excitation (labeled as **#**) are observed in both pristine and treated on-indent regions in good agreement with literature,[33,34] albeit with intensity variation due to spatial sampling mentioned above. In contrast, a new peak appears around 300 cm$^{-1}$ (labeled as **\***) in the voltage-treated regions, shown in more detail in the inset of Figure 2(**b**), which corresponds to the O-W-O bending mode indicative of oxidized WSe$_2$.[35–37] This oxidation-related peak is evident in both on-indent and off-indent voltage-treated regions, confirming that VIO alters the material structure beyond the indentation site. The persistence of the WSe$_2$ Raman peak in the on-indent region alongside the appearance of the oxidation peak supports the interpretation that pristine material is preserved within indentations, while surrounding regions undergo partial oxidation.

Furthermore, the Raman signal from the off-indent treated region (blue) retains a weak but discernible WSe$_2$ mode ($A^1_g$ + $E^2_g$) at ~250 cm$^{-1}$, as shown in the inset of Figure 2(**b**). The significant reduction in intensity by over two orders of magnitude indicates that the oxidation process is not fully complete in these regions. However, this residual signal does not compromise the overall effect of VIO, as the absence of corresponding PL emission in Figure 2(**a**) confirms that the material transformation (oxidation) is sufficient to suppress defect-bound exciton emission. This validates the effectiveness of VIO in enhancing the optical purity of QEs.

To further demonstrate the significant purity enhancement of monolayer WSe$_2$ QEs provided by the VIO method, we compare low-temperature PL and second-order correlation function $g^2(t)$ measurements of untreated and treated emitter areas. Figure 3 presents data from two representative emitter sites without and with the VIO treatment. These measurements, performed at 18K using a continuous-wave 705 nm laser, provide critical insight into the purity enhancement achieved through voltage-induced oxidation treatment. Figure 3(**a**) presents a typical PL spectrum of untreated monolayer $WSe_2$ quantum emitters, characterized by multiple sharp and narrow emission features attributed to nanoindentation-induced QEs.[18] As a comparison, PL spectra from bare monolayer $WSe_2$ without indentation, shown in Figure S4(**a**), exhibit only a broad background emission with no dominant sharp peak indicating the absence of QEs and the presence of defect-bound exciton luminescence.

The PL peak at ~736 nm in Figure 3(**a**) was selected for $g^2(t)$ measurements using 4 nm bandpass filter (yellow bar). A value of $g^2(0)$ = 0.456 ± 0.058 was obtained. Several factors limit the achievable QE purity in untreated samples: (1) overlapping PL peaks with broad background emission from defect-bound excitons due to material quality, and (2) multiple closely spaced emission peaks that complicate spectral isolation. Furthermore, the laser spot size ~2 µm significantly exceeds the indentation size (~300-600 nm), resulting in unwanted collection of background emission that degrades the measured single-photon purity. These confounding effects collectively lower the apparent single-photon purity and introduce substantial measurement uncertainty, which VIO aims to address.

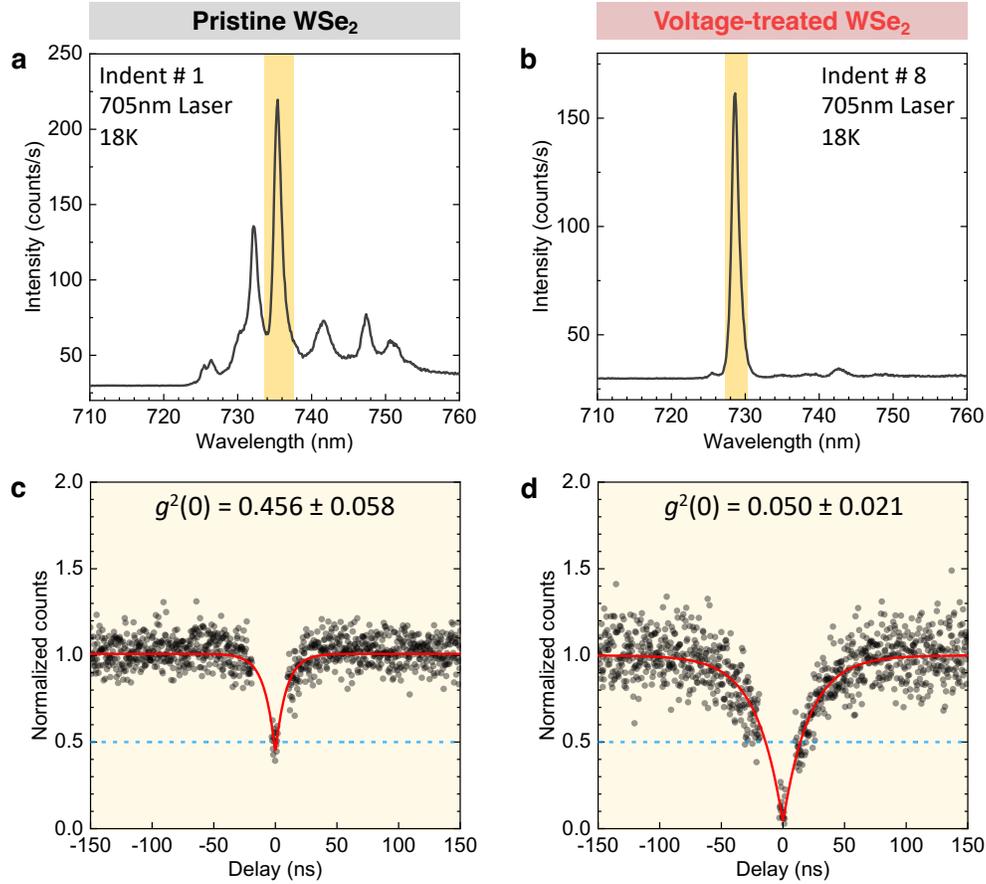

**Figure 3. Enhancement of single-photon purity in monolayer WSe$_2$ quantum emitters via VIO treatment.** (**a,b**) Photoluminescence (PL) spectra of representative quantum emitters (QEs) obtained from pristine (a) and voltage-treated (b) regions. The emitter in the pristine sample (a) shows multiple narrow peaks overlaid on a broad defect-bound exciton background, while the emitter in the treated sample (b) displays a single dominant emission peak with strongly suppressed background. (**c,d**) Corresponding second-order autocorrelation measurements $g^2(t)$ from the highlighted PL features (~4 nm spectral window, shown in yellow) in (a) and (b), respectively. The QE from the pristine sample exhibits a $g^2(0)$ value of 0.456 ± 0.058 in (c), indicating limited purity due to spectral overlap and background emission. The QE from the voltage-treated region (d) shows $g^2(0)$ = 0.050 ± 0.021, confirming high single-photon purity well below the $g^2(0) < 0.5$ threshold. Note that, due to the instrumentation response, we observed backflash effects for all the $g^2(t)$ measurements due to the reemission from detectors, which are further discussed in Figure S6. Lifetime $\tau$ is measured to be 8.79 ± 0.61 ns and 21.47 ± 0.70 ns for the emitters shown in (c) and (d), respectively. All measurements were conducted at 18 K using a CW 705 nm laser.

The impact of the VIO treatment is clearly evident in Figure 3(**b**), where the PL spectrum is dramatically simplified to a single sharp emission peak with negligible background intensity. This spectral purification directly improves the quality of quantum emission by eliminating the issues associated with overlapping emission features and minimizing contributions from defect-bound excitons. The resulting single-emitter dominance also improves $g^2(t)$ measurement fidelity by reducing ambiguity in photon correlation data.

This enhanced spectral isolation significantly boosts single-photon purity, as confirmed by correlation measurements shown in Figure 3(**c**) and 3(**d**). A representative untreated emitter exhibits a $g^2(0)$ value of 0.45 (QE Purity = 55%, Figure 3(**c**)), indicating considerable multiphoton contamination due to background noise and indistinguishable QEs. In stark contrast, an emitter that has been subjected to VIO treatment achieves a $g^2(0)$ value of approximately 0.05 (QE Purity: 95%, Figure 3(**d**)), clearly demonstrating the suppression of spurious emission from background and the realization of high-purity single-photon generation. This significant improvement in purity from 55% to 95% surpasses the widely accepted threshold of $g^2(0) < 0.5$ for single-photon sources, underscoring the effectiveness of VIO. Moreover, by isolating a single quantum state and reducing spectral noise, VIO contributes to stable and reliable emitter performance, critical for scalable integration into quantum photonic technologies.

Figure 4 further reinforces the effectiveness of VIO treatment by presenting statistical data across 14 randomly selected quantum emitters: 7 without, and 7 with voltage treatment. The results reveal a consistent trend: prior to VIO, most emitters exhibited $g^2(0)$ values at or above 0.5, indicating that the emitted light was dominated by classical emission. In contrast, following VIO treatment, the majority of emitters achieved $g^2(0)$ values well below 0.5, confirming quantum emission with high purity.

The improvements in PL and raw $g^2(0)$ plots are shown in Figure S5. These observations highlight the robustness and scalability of the VIO process as an effective technique for improving the performance of quantum emitters. The VIO method introduced in this study effectively addresses the purity limitations observed in untreated samples. By applying a controlled voltage to indented monolayer $WSe_2$, VIO globally suppresses defect-bound exciton emission while preserving QE regions within the indentation. This selective oxidation eliminates spectral overlap from indistinguishable QEs and minimizes background luminescence.

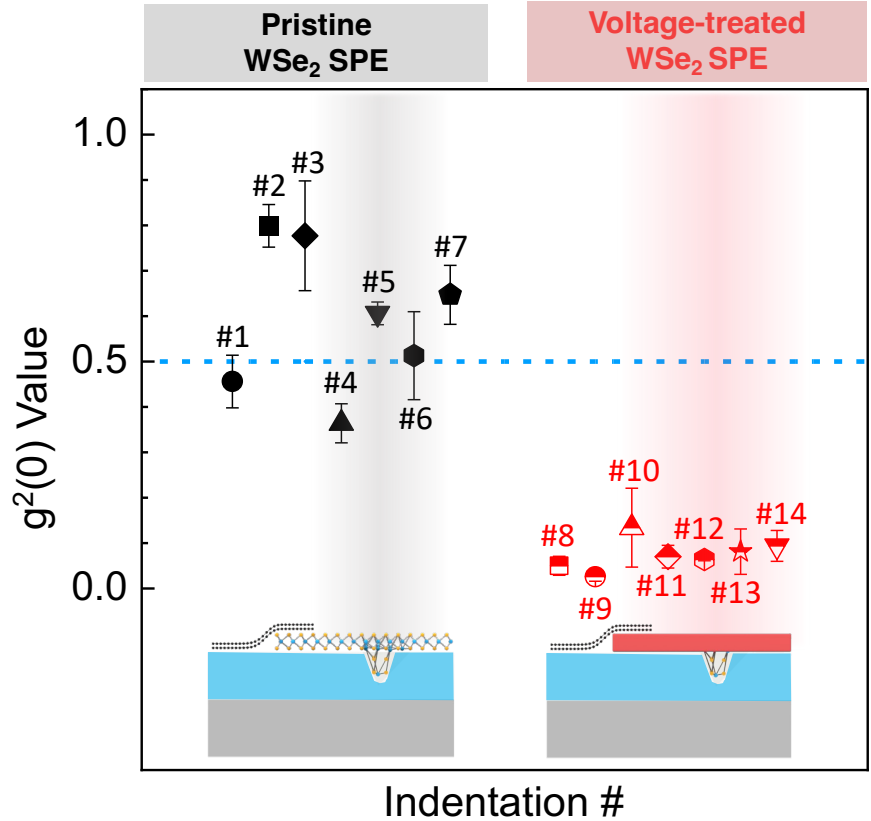

**Figure 4. Summary of second-order correlation $g^2(0)$ values for pristine (black) and voltage-treated (red) WSe$_2$ QEs from 14 randomly selected indentation sites.** Emitters from pristine regions exhibit a wide range of $g^2(0)$ values, with many near or above the 0.5 threshold (blue dashed line), indicating poor single-photon purity due to background emission and indistinguishable emitters. In contrast, voltage-treated emitters consistently displayed $g^2(0)$ values well below 0.5, confirming high-purity single photon emission enabled by the voltage-induced oxidation (VIO). All PL and $g^2(t)$ associated with these emitters are presented in Figure S5.

We note that a direct one-to-one comparison of the same indentation site before and after VIO treatment was not performed in this study due to the irreversible nature of the oxidation process. Instead, we evaluated two statistically sampled groups of emitters; one untreated, and one voltage-treated. While post-treatment re-measurement of the same sites was technically feasible, this approach was avoided to prevent complications from irreversible site modification, such as changes in the local strain profile due to thermal cycling or deformation induced during the applied-bias oxidation process. This precaution ensures the integrity of the emitter characteristics and does not affect the validity of our conclusions, as the statistical differences in $g^2(0)$ values between the two groups were consistent and significant.

Moreover, to evaluate the broader yield and performance of SPEs, we further analyzed more than 58 indentation sites using the intensity signal-to-noise (S/N) ratio ($I_{SPE}/I_{background}$) shown in Figure S7 and Table S1. While $g^2(0)$ values are not directly included in Table S1, the S/N ratio serves as an empirical indicator of purity, supported by the clear inverse correlation observed in the 14 emitters analyzed in Figure S5. Sites with higher intensity ratio consistently demonstrated lower $g^2(0)$, indicating enhanced single-photon emission characteristics. Although we do not claim a strict quantitative relationship between intensity ratio and $g^2(0)$, the qualitative trend is supported by earlier works[21–23] and offers a scalable metric for rapid and efficient pre-screening emitter candidates. Table S1 thus provides an extended statistical assessment of emitter yield and quality enabled by voltage-induced oxidation.

Figure 3 and Figure 4 collectively demonstrate that voltage-induced oxidation treatment significantly improves both the purity and reproducibility of single-photon emitters in monolayer $WSe_2$. This enhancement is attributed to the selective suppression of defect-bound excitonic emissions surrounding the quantum emitter sites, which suppresses defect-bound exciton emission and creates a lateral heterostructure - defined here as an in-plane interface between pristine (emissive) and oxidized (non-emissive) regions within the same monolayer. As a result, VIO provides a rapid and efficient strategy for increasing both the yield and optical quality of 2D quantum emitters, paving the way for scalable integration into quantum photonic platforms.

In conclusion, a major challenge in $WSe_2$ quantum emitter research has been the presence of multiple emission peaks with closely spaced energies and significant defect-bound exciton contributions, which collectively degrade the purity and reproducibility of quantum emission. Our voltage-induced oxidation (VIO) technique addresses these challenges by selectively oxidizing the surrounding $WSe_2$ material, thereby suppressing background excitonic emissions and preserving pristine quantum emitters within indentation sites. This process leads to a simplified emission spectrum, dramatically improved single-photon purity, and enhanced emitter yield. Importantly, this method operates in a nonvolatile manner, requiring no ongoing external fields or complex fabrication beyond standard nanoindentation and c-AFM treatment. While voltage-induced oxidation (VIO) improves emitter purity and reproducibility, precise control over the emission wavelength of $WSe_2$ quantum emitters remains an open challenge. Future work should focus on incorporating spectral tuning strategies such as strain engineering, local dielectric patterning, or electrostatic gating in conjunction with VIO. Such advances will enable the deterministic and scalable quantum light sources with tailored emission characteristics, accelerating their integration into quantum photonic circuits and next-generation quantum technologies.

## MATERIALS AND METHODS:

**Material growth of monolayer $WSe_2$:**

Quartz tube furnaces are used to synthesize monolayer $WSe_2$ on $SiO_2$/Si substrates (275 nm oxide) following previously outlined procedures.[38] Perylene-3,4,9,10-tetracarboxylic acid tetrapotassium salt molecules are deposited on the growth substrate to promote lateral growth. High-purity metal oxide and chalcogen powder serve as precursors for the growth and ultra-high purity argon flows continuously through the furnace as it heats to the target temperature of approximately 850°C. Upon reaching the target temperature, ultra-high purity $H_2$ is added to the Ar flow and maintained throughout the desired soak time and subsequent cooling to room temperature.

**Sample fabrication:**

All samples were carefully fabricated on the P(VDF-TrFE)-Si substrate. The desired thickness (~260 nm) of P(VDF-TrFE) was transferred onto clean Si substrate followed by transferring CVD-grown monolayer $WSe_2$ flakes and thin graphite flake on the side contact to the monolayer $WSe_2$ flakes. AFM nano-indentation is then performed to create indention array on monolayer $WSe_2$ flake for QEs generation. All detailed steps of the sample fabrication can be found in our previous study.[22]

**Voltage treatment by conductive AFM:**

For the voltage treatment by AFM, the Park AFM NX-20 system is used under PFM mode with Multi75-E, Cr/Pt-coated Si cantilever (PFM cantilever) and applying direct current (d.c.) voltages ranging from +10 V to -10 V to the AFM tip and substrate, $V_{tip}$ and $V_s$, respectively (total bias applied was 20 V.) The AFM tip is scanning across the graphite/$WSe_2$ area (5 µm by 1.5 µm with 0.5 Hz scan rate) while applying the bias to the AFM tip. The 10 nN is also applied to the AFM tip to the substrate providing gentle treatment in PFM mode (contact mode). The constant bias of -10 V was first applied to $V_s$ while $V_{tip}$ remains 0 V and then slowly increase the $V_{tip}$ from +0.01 V to +10 V in 1 min. Keep constant $V_{tip}$ = 10 V until the optical contrast of the $WSe_2$ changes (6 min in total for the sample shown in the main text.). The gradual change in optical color contrast of $WSe_2$ can be observed from the AFM camera showing examples in Figure S1.

**Fluorescence imaging, Photo L , and Raman for initial assessment:**

The fluorescence(FL) images presented in Figure 2 was obtained using an Olympus BX53M microscope with a 100X objective lens. Illumination was provided by an X-Cite Xylis XT720S LED fluorescence light system, in conjunction with a wideband green fluorescence filter cube (Olympus U-MWG2).

Room temperature photoluminescence(PL) and Raman measurements were performed in a scanning confocal microscope (Horiba LabRam Evolution) using 532 nm continuous-wave laser and 50X objective to a spot size of 2 µm. The corresponding power density for room temperature PL are ~5 µW µm$^{-2}$ for pristine $WSe_2$, ~250 µW µm$^{-2}$ for voltage treated $WSe_2$ (oxidized $WSe_2$). And the corresponding power density for Room temperature Raman are ~16 µW µm$^{-2}$ for both pristine and voltage treated $WSe_2$ (oxidized $WSe_2$). 600 l/mm and 1800 l/mm grating were selected for PL and Raman respectively. PL data for oxidized $WSe_2$ was acquired with longer accumulation time and higher laser power due to the strong suppression of the PL intensity. All data were converted to counts/sec for better comparison.

Low temperature PL data in Figure S7, (**b**) and (**c**) was acquired by the same scanning confocal microscope (Horiba LabRam Evolution) in conjunction with the Montana Cryostation C2 system with 532 nm continuous-wave laser and 50 ULX (ultralong walking distance) objective to a spot size of 2 µm. The laser power density at sample is ~3.20 µW µm$^{-2}$.

**Low temperature PL and second order correlation measurements by PicoQuant:**
Single-photon emission was verified by measuring the second order correlation measurements, g$^2$(t), in Hanbury–Brown Twiss geometry. Measurements were conducted using a PicoQuant MicroTime 200 time-resolved confocal microscopy platform equipped with a 50:50 beam splitter, two time-tagged single-photon avalanche photodiodes (Excelitas SPCM-AQRH-14) having detector efficiencies above 65% in the spectral range from 650 to 750 nm, and time-correlated single photon counting electronics (PicoQuant HydraHarp 400). The sample was cooled to 18 K in a closed-cycle Lakeshore ST-500 cryostat. Piezo motor-driven stages move an ex-situ 100x glass-corrected objective (Olympus LCPLFLN100XLCD) relative to the stationary sample and cryostat, allowing for sample mapping. The 705 nm excitation laser is blocked using a combination of a 705 nm dichroic beamsplitter and 705 nm longpass filter. Short pass filters (Thorlabs FESH0750) were positioned before each detector to reduce the collection of unwanted backflash emission produced by the photodiodes. The system backflash cannot be fully removed and leads to strongly correlated features in the g$^2$(t) data at approximately ±6 ns. These peaks are removed from the presented figures for simplicity, and discussed further in the supplementary information. To reduce the collection of classical light, a bandpass filter (Spectrolight FWS Mono 745) is tuned to match the emitter wavelength for each SPE. Photon correlation histograms were generated using 300 ps time bins. Values of g$^{(2)}$(0) and lifetime $\tau$ were extracted by fitting to
$$g^2(t) = 1 - \rho^2((1+a)e^{-\frac{|t-t_0|}{\tau_1}} - ae^{-\frac{|t-t_0|}{\tau_2}})$$ when a shelving state is present, or to
$$g^2(t) = \left(\frac{n-1}{n}\right) + \frac{1}{n}(1 - e^{-\frac{|t-t_0|}{\tau_1}})$$ in the absence of a shelving state.

**Acknowledgements**

The research performed at the Naval Research Laboratory was supported by core programs. This research was performed while S.-J.L. held an American Society for Engineering Education fellowship at NRL.

**Contributions**

S.-J.L., H.-J.C., and B.T.J. conceived and designed the experiments. S.-J.L., H.-J.C., K.M.M., fabricated the samples, and characterized with PL, Raman, AFM, and PFM. S.-J.L., K.M.M., and M.A.N., performed PL and the time correlated single photon counting measurements. S.-J.L., H.-J.C., wrote the manuscript. All authors discussed the results and commented on the manuscript.

**Ethics declarations**

The authors declare no competing interests.